\documentstyle[12pt]{article}
\begin{document}

\title{The fundamental plane of galaxy clusters}

\author{R. Schaeffer$^{1}$, S. Maurogordato$^{2}$, A. Cappi$^{3,4}$
and F. Bernardeau$^{1}$ \\
\small $^{1}$ Service de Physique Th\'eorique, CE Saclay, 91191 Gif sur Yvette,
France \\
\small $^{2}$CNRS UA 173; DAEC, Observatoire de Paris - Meudon, 92195 Meudon,
France \\
\small $^{3}$Universit\'e Paris VII; DAEC, Observatoire de Paris - Meudon,
92195 Meudon France \\
\small $^{4}$Osservatorio Astronomico di Bologna, Via Zamboni 33,
40126 Bologna, Italy}

\maketitle

\begin{abstract}
Velocity dispersion $\sigma$, radius $R$ and luminosity $L$ of
elliptical galaxies are known to be related,
leaving only two degrees of freedom and defining the so-called
``fundamental plane". In this {\em Letter} we present
observational evidence that rich galaxy clusters exhibit a similar behaviour.
Assuming a relation $L \propto R^{\alpha}\sigma^{2 \beta}$,
the best-fit values of $\alpha$ and $\beta$ are
very close to those defined by galaxies.
The dispersion of this relation is lower than 10 percent, i.e.
significantly smaller than the dispersion observed in the
$L-\sigma$ and $L-R$ relations.
We briefly suggest some possible implications on the spread of formation times
of objects and on peculiar velocities of galaxy clusters.
\end{abstract}

\baselineskip 18pt

\section{Introduction}

The distribution of structures in the Universe according to mass and size is
believed to be intimately related to the conditions prevailing in the
primordial Universe, and reflects the history of formation.
The existence of a universal luminosity function (Schechter, 1976), backed by
a dynamical model (Press \& Schechter, 1974) of clustering, led to the belief
that the luminosity (or the mass) is the major, if not the only, parameter
describing a galaxy.
Other properties such as the velocity dispersion
(Faber \& Jackson, 1976; Tully \& Fischer, 1977) are then related
to the luminosity (Cavaliere et al., 1978) and provide important distance
indicators.
Analogue relations exist for smaller systems:
for dwarf ellipticals  Held et al. (1992) have found $L \propto \sigma^{2.5}$,
for globular clusters Meylan \& Mayor (1986) and
Paturel \& Garnier (1992) have found $ L \propto \sigma^{2.5}$.
Soon, however, the need for at least a second parameter
for spiral galaxies (Bujarrabal et al., 1981)
as well as for ellipticals (Terlevich et al., 1981, Tonry \& Davis, 1981)
became apparent. Subsequently,
elliptical galaxies were shown to genuinely form a two-parameter family,
the so-called (Djorgovski \& Davis, 1987; Dressler et al., 1987)
``fundamental plane", and the second parameter could be identified
as being the surface brightness (or equivalently the radius).
Nieto et al. (1990) have found that low-mass ellipticals, dwarf
spheroidal galaxies and halo globular clusters
are all near the plane defined by bright elliptical galaxies.
Despite the identification of the relevant parameters, the reason for
they being  two is not known, except for an
obvious relation with the virial theorem, which can be written in the form
$ M  \propto R \sigma^2  $.
Internal dynamics in pressure supported systems, dissipative merging,
cooling, and star formation
processes were invoked by various authors (Dressler et al., 1987;
 Nieto et al., 1990;
Kashlinsky, 1983) having in mind that this relation applies to extremely dense
systems.
A general discussion can be found in Kormendy \& Djorgovski (1989),
and references therein.

In this {\em Letter} we show that:

a) the fundamental plane extends to large bound systems of much lower density
such as clusters of galaxies;

b) the fact that systems as small as globular clusters
and as large as galaxy clusters lie on a fundamental plane,
suggests that stellar systems as galaxy systems
were formed following the same process, thus supporting
the hierarchical clustering scenario;

c) the dispersion seen in the luminosity - radius or luminosity - velocity
dispersion relations should reflect the dispersion in the epoch of formation.

\section{Discussion of data}

\subsection{Abell clusters}

In order to search for correlations between intrinsic parameters of galaxy
clusters, it is essential to work on an homogeneous data set. Therefore we
have used the compilation of West, Oemler \& Dekel (1989; herafter WOD)
of 29 Abell clusters with reliable photometry.
For a more exhaustive discussion of this sample, we address the reader to
their paper (and also to West, 1990).
High-quality profiles are derived (West, Dekel \& Oemler, 1988) allowing good
fits by a de Vaucouleurs law and so determination of an accurate
half-mass radius $R_e$.
The photometry of these  clusters, previously
observed by Oemler (1974), Butcher, Oemler \& Wells (1983) and Dressler (1978),
has been re-reduced extrapolating the previously calculated
profiles to infinite radius, and using the luminosity function of Kirshner
et al. (1983) for extrapolation to faint galaxies where
$L$ was given in units of
$L_* = 1.3 \ 10^{10} \ L_{\odot} $ in the V passband (with $H_0 = 100$ km/s).
Velocity dispersions are derived from the compilation of Struble \& Rood (1991)
and allow us to build a sub-sample of 16 clusters with reliable
effective radius $R_e$, total luminosity $L$ and velocity dispersion $\sigma$:
A154, A168, A194, A400, A401,
A426, A539, A665, A1314, A1656, A1904, A2019, A2065, A2199, A2256, A2670.
$\sigma$ has been measured with more than 50 galaxies
per cluster for 75 \% of the sample and between 13 and 50 galaxies per cluster
for the remaining 25 \%.
These clusters are rich and essentially free from superposition effects,
so that the uncertainty in $\sigma$ is not critical.
The values of $ \chi^2$ for this sample have been calculated
using errors of 15\% for each variable (WOD).

\subsection{Stellar systems}

Data for ellipticals are taken from Djorgovski \& Davis (1987).
We have converted their Lick $r_G$ magnitudes (measured within $R_e$
to the V passband using Djorgovski's (1985) approximate relations and assuming
$B-V = 1$; we adopt a luminosity two times that in $R_e$.
The second sample of ellipticals used is that of Faber et al. (1989).
We have used V magnitudes from the listed total B magnitudes and B-V colour
when available; otherwise B-V = 1 has been assumed.
Being interested in a general comparison
between elliptical galaxies and galaxy clusters, we will
will ignore more detailed corrections.
We have also added a sample of low-luminosity
and true dwarf elliptical galaxies, but uncertainties here are quite large.
Data for low luminosity ellipsoidal galaxies are taken
from Bender \& Nieto (1990), and for true dwarf elliptical galaxies from
Bender et al. (1991); we assume $B-V = 0.7$ (as in Held et al., 1992).

Illingworth (1976) has studied 10 clusters, fitting them both with a King
and a de Vaucouleurs law, and listing their
effective radius, total visual luminosity,
central velocity dispersion and mass.
The central velocity dispersions of other clusters have been
taken from Gunn \& Griffin (1979),
Meylan \& Mayor (1986), Lupton, Gunn \& Griffin (1987),
Peterson \& Latham (1986), Pryor et al. (1988),
Rastorguev \& Samus (1991) and Zaggia et al. (1991).
All these authors except Illingworth and Zaggia et al.
calculated velocity dispersions from radial velocities of giant stars in
globular clusters (note that dispersions from integrated spectra
appear to be systematically higher than those calculated from radial
velocities of giant stars).
Except for Illingworth's clusters, we have only
core and tidal radii listed by Webbink (1985), who gives also
integrated V absolute magnitudes (within $r_t$), and not de Vaucouleurs radii.
Being interested in {\em global} quantities, we have estimated $R_e$ as a
function of $r_c$ and the concentration parameter $c = log(r_t/r_c)$,
where $r_t$ is the tidal radius.
Our final sample includes 33 globular clusters (13 with velocity
dispersions derived from integrated spectra).

\section{Discussion of results}

\subsection{Luminosity - Radius and Luminosity - velocity dispersion relations}

WOD have shown the existence of a correlation between the
total luminosity and the effective radius of Abell clusters.
We get $ R \propto L^{0.5 \pm0.1}$, but with a high dispersion
($\chi^2 = 2.7$ per degree of freedom), in accordance with WOD.
Fitting the inverse relation to the same data gives
$ L \propto R^{1.34 \pm 0.17}$ with $\chi^2 \sim 3.2$ per DOF
(fig.1).

We note that the two fits are in principle not compatible. This is to be
expected in a fit when the measurement errors are much smaller than the
actual dispersion of the points, and just reflects the large value of the
$\chi^2$.

A correlation between richness and velocity dispersion
for Abell clusters is known to exist (Danese et al., 1980; Cole, 1989).
In the subsample we have extracted from WOD,
a relation between total luminosity and  velocity dispersion
has been seeked. We get $ L \propto \sigma^{1.87 \pm 0.44} $,
but (fig.2) with a dispersion
($\chi^2 = 1.9$ per DOF) uncomfortably large for this relation to be
statistically acceptable.  Such deviations have often been interpreted as being
the result of error underestimates due to the existence of sub-clustering
or  to a possible erroneous identification of cluster members.

\subsection{The fundamental plane for clusters}

Seeking, on the other hand, for a relation $L \propto R^\alpha\sigma^{2\beta}$,
we get $\alpha = 0.89 \pm 0.15 $ and $\beta = 0.64 \pm 0.11$, with
(fig.3)
a considerable improvement ($\chi^2 = 0.38$ per DOF).

{\em We must emphasize that the evidence of the galaxy clusters fundamental
plane relies on the careful photometric work done by
WOD}. Would we use richness as an approximate luminosity,
the correlation would be nearly lost. This is why such a relation had
not been noticed earlier. The key of our following argumentation
is the low relative value of $\chi^2$ for
the fit. It shows that the actual errors on the determination of $L$, $R_e$ and
$\sigma$ were by no mean larger than the estimation we used, but simply that
Abell clusters form a two-parameter family following a very tight relation.
If for instance, we remove from our sample the four clusters belonging to
well-known superclusters as Coma, A2199, Corona Borealis and Perseus, and
A194 which is elongated and A401 which may interact with A399,
the values of $\alpha$ (0.90) and $\beta$ (0.66)
are found to be within the quoted errors,
with a stable $\chi ^2$. Subclustering or intruders can give errors
that are at most at the $10\%$ level.
The values of $\alpha$ and $\beta$ we find for clusters
are close to those obtained in the litterature for the galaxies:
$\alpha = 0.75$, $\beta = 0.83$ by Dressler et al. (1987),
$\alpha = 0.92$, $\beta = 0.93$ or
$\alpha = 0.89$, $\beta = 0.77$ by Djorgovski and Davis (1987),
with uncertainties comparable to ours, as well as those we have
determined using the sample of 15 dwarf and low-luminosity ellipsoidal
galaxies, $\alpha = 0.85 $, $\beta = 1.$, and those for
33 globular clusters,
$\alpha = 0.68 \pm 0.12$, $\beta = 0.71 \pm 0.05$ (a fit to the globular
clusters with dispersions calculated from individual stars gives
$\alpha = 0.70$ and $\beta = 0.73$).
In this sense our conclusion is that
a fundamental plane exists for
{\em matter condensations with extremely
different masses, i.e. globular clusters, galaxies and
rich clusters} (fig.4).

As $\alpha$ and $\beta$ are not very different, we fitted the
relation $L = K (R \sigma^2)^\gamma$.
This allows to determine directly the constant $K$ which
caracterizes the gap between the planes,
and to find the average $\langle M/L \rangle$ ratio between the different
classes of objects.
The resulting values are summarized in table 1 which shows the
stability of the slope  and the consistence of the $\langle M/L \rangle$ ratios
with previous estimations of Oemler (1974) and Illingworth (1976).

\subsection{Implications on structure formation}

A natural outcome of structure formation models
(Schaeffer \& Silk, 1985, 1988) based
on hierarchical clustering is that all mass condensations form
a two-parameter family (see also Peacock \& Heavens, 1985,
Kaiser, 1988, Peacock, 1990, who addressed the same question).
Primordial mass fluctuations in an Einstein-de~Sitter
Universe, $\delta(M) = \delta_0(M) (1+z)^{-1}$ at scale
$M$ and epoch $t$ --related  to the redshift by $t \propto (1+z)^{-3/2}$--
are believed to follow a gaussian random distribution with
$ \langle \delta_0^2(M) \rangle ^{1/2} = \Sigma(M)$. A given fluctuation
collapses
when $\delta(M) \sim 1 $, at an epoch $(1 + z_{form}) \sim \delta_0(M)$
to an object with a final density
$\rho \propto (1+z_{form})^3$ $\propto$ $\langle \delta_0^3(M) \rangle$
that is proportional to (Gott \& Rees, 1975; Peebles, 1980)
the density of the Universe at the formation time.
This implies $R \propto M^{1/3} / \delta_0 (M)$ and
$\sigma^2 \propto M^{2/3} \delta_0(M)$. Due to the fluctuations of $\delta_0$,
both $R$ and $\sigma$ fluctuate and are not simply a function of mass.
The product $R \sigma^2 \propto M$ however is independent of $\delta_0$.
The relation between mass and luminosity, that originates from the
intrinsic physical processes that govern star formation, too, has a very
low dispersion. This implies $M/L \propto L^\epsilon$, where
$\epsilon = 2/(\alpha+\beta)-1$ or
$ L \propto (R\sigma^2)^{\frac{1}{1+\epsilon}} $,
and is consistent with the previous findings provided the
difference $\alpha - \beta$ is compatible with zero.
Indeed, we have $\alpha - \beta$ = 0.2 $\pm$ 0.2 (Abell clusters),
-0.1, 0.0 or 0.1 (galaxies, Dressler et al., 1987, Djorgovski \& Davis, 1987),
-0.1  (dwarf galaxies, Bender \& Nieto, 1990) and $0.1 \pm 0.2$
(globular clusters).
The mass is found to vary nearly as the
luminosity does: $\epsilon = 0.3 \pm 0.1$ (Abell clusters),
0.3, 0.1, 0.2 (galaxies), -0.1 (dwarf galaxies),
$0.4 \pm 0.1$ (globular clusters) respectively. Except for dwarf galaxies,
the mass has a slight tendency to increase faster than the luminosity does. To
show that $M/L$ deviates significantly from a constant value would require
further work.
The point here is simply that $M$ is much more tightly correlated to $L$ than
$R$ or $\sigma$ are:
{\em the dispersion seen in the luminosity - radius or luminosity - velocity
dispersion relations should reflect the dispersion in the epoch of formation}.
Indeed, the values of $\chi^2$ can be used to obtain the dispersion of
$\delta_0$, whence of $z_{form}$:
$ \Delta_z \equiv
( \langle (1+z_{form})^2 \rangle - \langle 1+z_{form}\rangle ^2 ) ^{1/2}
/ \langle 1+z_{form} \rangle$.
The calculation can be schematically summarized as follows. We assume
that the dispersion in radius is due to random observational errors
with dispersion $\Delta_{obs}$ as well as to a random redshift of
formation with dispersion $\Delta_z$ and link the latter to the observed
dispersion in excess of the expected one. This procedure is obviously sensitive
to the adopted observational errors that are usually not determined with
an extremely high accuracy.
We find $\Delta_z \sim 0.2$-$0.3$ for Abell clusters and
$\Delta_z \sim 0.4$-$0.6$ for elliptical galaxies.

A theoretical formulation (Schaeffer \& Silk, 1985) based on hierarchical
clustering, along the lines developed by Press and Schechter, but with
the difference in formation times explicitely taken into account, leads to
$ \Delta_z = \left( \Sigma(M) / \delta_c \right) ^2 \sim 0.1$, with
$\delta_c \sim 1.7$,
for rich clusters ($\Sigma(M) \sim 0.6$), a value that increases for
the smaller mass objects
to $ \Delta_z = \sqrt{\pi/2 - 1} = 0.76 $ at scales where
the fluctuation $\Sigma(M)$ are large, as is expected for galaxies.
These theoretical results are comparable to the observed values.
Anyway more work is required to take into account not only
observational errors, but also the effect of the distinct dynamical processes
which led to the formation of galaxy clusters, galaxies and globular clusters.

Moreover, if we assume that the distance of each cluster from the fit of
the fundamantal plane is entirely due to his peculiar motion, it follows
that most of the cluster peculiar velocities relative to the
CMB are less than 1000 km/s.

\section{Conclusions}

We have shown that galaxy clusters lie in the fundamental plane.
This common property of stellar systems and
galaxy clusters suggests a similar process of
formation, favouring the hierarchical clustering scenario;
the dispersions in the observed relations $L - R$ and
$L - \sigma$ should reflect the dispersion in the epoch of formation.

The small dispersion (8\%) could allow a direct estimation of
distances for galaxy clusters giving access to their large scale motions.
Moreover, the existence of the FP gives us
further information of the distribution of luminous and dark components
of matter in clusters. We will analyse these subjects in future work.

Accurate measurements on a larger sample are strongly needed to have
better estimates of the relevant parameters; especially a careful
determination of the measurement errors is important.
An observational program is in progress in order to test the
relation on ten more clusters.

\section*{Acknowledgement}

{We wish to thank A.Dekel for useful discussions.}

\newpage

%
%
\begin{table}
\centering
      \caption{Best fit values for the relation $L = K (R \sigma^2)^\gamma$
                  and relative $\langle M/L \rangle$ ratios.}
       \label{tab:par}
	\begin{tabular}{lccc}
                    & $\gamma$ & $K$	& $\langle M/L \rangle /
\langle M/L \rangle _{gal}$ \\
Abell clusters 	  & 0.73 & 1.1 $10^8$ &	~40 \\
ellipticals $^1$  & 0.78 & 3.8 $10^8$ &	~~1 \\
ellipticals $^2$  & 0.82 & 3.0 $10^8$ &	~~1 \\
globular clusters & 0.70 & 7.4 $10^7$ &	~~$0.1(H_0/100)^{-1}$ \\
            \end{tabular}
{$^1$} Faber et al., 1989; {$^2$} DD87.
            \end{table}

\newpage

\bigskip

{\bf Figure 1:} Luminosity -- radius relation for the compilation
(West et al., 1989) of 29 Abell clusters
($R$ is the effective radius in Mpc, assuming $H_0$ = 100 km s$^{-1}$
Mpc$^{-1}$).
The stars represent the 16 clusters for which
the velocity dispersion is known. The fits for an $R \propto L^\delta$
relation with $\delta = 0.5 \pm 0.1$ as in ref. (West et al., 1989)
(dashed-dotted line),
and for an $L \propto R^\alpha$ relation with $\alpha = 1.34 \pm 0.17$
(full line) are shown.
Due to the  dispersion of the points that is much larger than the measurement
errors, the two procedures
are not equivalent.

{\bf Figure 2:} Luminosity -- velocity dispersion relation,
with a fit (full line) for $L \propto \sigma^{2\beta}$, $\beta = 0.94$.

{\bf Figure 3:}
Relation between luminosity L and the product $R^\alpha\sigma^{2\beta}$.
Note the excellent fit ($\alpha=0.89$, $\beta=0.64$ with a constant factor
$ K=4.10^8$), for which the $\chi^2$
per degree of freedom is improved by a factor of 8 as compared to the previous
cases.

{\bf Figure 4:}
The ``fundamental plane" seen edge-on for different systems.
Crossed circles: globular clusters with individual stars spectra,
triangles: globulars clusters with integrated spectra,
squares: dwarf and low-luminosity
ellipsoidal galaxies (Bender \& Nieto, 1990; Bender et al., 1991),
crosses: elliptical galaxies (Djorgovski \& Davis 1987),
circles:  elliptical galaxies (Faber et al. 1989)
stars: galaxy clusters (West et al., 1989; Struble \& Rood, 1991).

\end{document}